\documentclass[conference]{IEEEtran} %

\usepackage{balance}
\usepackage[utf8]{inputenc}

\IEEEoverridecommandlockouts 
\usepackage[dvips]{graphicx}
\usepackage{algorithmic}
\usepackage{algorithm}
\usepackage{listings}
\usepackage[cmex10]{amsmath}
\usepackage{url}
\usepackage{multirow}
\usepackage{pgfplots}

\usepackage{amssymb}
\usepackage{amsthm}
\newtheorem{definition}{Definition}

\title{Centrality Measures in multi-layer Knowledge Graphs}
\author{
\IEEEauthorblockN{Jens D\"orpinghaus\IEEEauthorrefmark{1}%
, Vera Weil\IEEEauthorrefmark{3}, Carsten Düing\IEEEauthorrefmark{2}, Martin W. Sommer\IEEEauthorrefmark{4}}%
\IEEEauthorblockA{%
\IEEEauthorrefmark{1} Federal Institute for Vocational Education and Training (BIBB), Bonn, Germany,\\ Email: jens.doerpinghaus@bibb.de, \url{https://orcid.org/0000-0003-0245-7752}}
\IEEEauthorblockA{%
\IEEEauthorrefmark{2} Mathematical Institute, University Koblenz-Landau, Koblenz, Germany,\\ Email: dueing@uni-koblenz.de}
\IEEEauthorrefmark{3} Department for Computer Science, University of Cologne, Germany
\IEEEauthorrefmark{4} Argelander-Institut für Astronomie, Bonn, Germany}

\newtheorem{dfn}{Definition}[section]

\newtheorem{exl}[dfn]{Example}

\begin{document}
\maketitle              %

\begin{abstract}
Knowledge graphs play a central role for linking different data which leads to multiple layers. Thus, they are widely used in big data integration, especially for connecting data from different domains. Few studies have investigated the questions how multiple layers within graphs impact methods and algorithms developed for single-purpose networks, for example social networks. This manuscript investigates the impact of multiple layers on centrality measures compared to single-purpose graph. In particular, (a) we develop an experimental environment to (b) evaluate two different centrality measures – degree and betweenness centrality -- on random graphs inspired by social network analysis: small-world and scale-free networks. The presented approach (c) shows that the graph structures and topology has a great impact on its robustness for additional data stored. Although the experimental analysis of random graphs allows us to make some basic observations we will (d) make suggestions for additional research on particular graph structures that have a great impact on the stability of networks.
\end{abstract}

\section{Introduction}

Knowledge graphs have been shown to play an important role in recent knowledge mining and discovery, for example in the fields of digital humanities, life sciences or bioinformatics. They also include single purpose networks (like social networks), but mostly they contain also additional information and data, see for example \cite{suarez2021risks,berhan2019board,rollinger2014amicitia}. Thus, a knowledge graph can be seen as a multi-layer graph comprising different data layers, for example social data, spatial data, etc. In addition, scientists study network patterns and structures, for example paths, communities or other patterns within the data structure, see for example \cite{dorpinghaus2019knowledge}. Very few studies have investigated the questions how multiple layers within graphs impact methods and algorithms developed for single-purpose networks, see \cite{rossetti2021conformity}. This manuscript investigates the impact of a growing part of other layers on centrality measures in a single-purpose graph. In particular, we develop an experimental environment to evaluate two different centrality measures -- degree and betweenness centrality -- on random graphs inspired by social network analysis: small-world and scale-free networks. 

This paper is divided into five sections. The first section gives a brief overview of the state of the art and related work. The second section describes the preliminaries and background. We will in particular introduce knowledge graphs and centrality measures. In the third section, we present the experimental setting and the methods used for this evaluation. The fourth section is dedicated to experimental results and the evaluation. Our conclusions are drawn in the final section.

\section{Preliminaries}

The term \emph{knowledge graph} (sometimes also called a \emph{semantic network}) is not clearly defined, see \cite{Fensel2020}. In~\cite{ehrlinger2016towards},  several definitions are compared, but the only formal definition was related to RDF graphs which does not cover labeled property graphs. As another example, \cite{paulheim2017knowledge} gives a definition of knowledge graphs limited to the definition of important features. Knowledge graphs were introduced by Google in 2012, when the Google Knowledge Graph was published on the use of semantic knowledge in web search, see \url{https://blog.google/products/search/introducing-knowledge-graph-things-not/}. This is a representation of general knowledge in graph format. Knowledge graphs also play an important role in the Semantic Web and are also called semantic networks in this context.

Thus, a \emph{knowledge graph} is a systematic way to connect information and data to knowledge. It is thus a crucial concept on the way to generate knowledge and wisdom, to search within data, information and knowledge. Context is the most important topic to generate knowledge or even wisdom. Thus, connecting knowledge graphs with context is a crucial feature. 

\begin{definition}[Knowledge Graph]\label{def:kg}
We define a knowledge graph as graph $G=(E,R)$ with entities $e\in E=\{E_1,...,E_n\}$ coming from formal structures $E_i$ like ontologies. 
\end{definition}

The relations $r\in R$ can be ontology relations, thus in general we can say every ontology $E_i$ which is part of the data model is a subgraph of $G$ indicating $O\subseteq G$. In addition, we allow inter-ontology relations between two nodes $e_1, e_2$ with $e_1 \in E_1$, $e_2 \in E_2$ and $E_1 \neq E_2$. In more general terms, we define $R=\{R_1,...,R_n\}$ as a list of either inter-ontology or inner-ontology relations. Both $E$ as well as $R$ are finite discrete spaces.

Every entity $e\in E$ may have some additional metainformation which needs to be defined with respect to the application of the knowledge graph. For instance, there may be several node sets (some ontologies, some actors (like employees or stakeholders, for example), locations, ...) $E_{1},...,E_{n}$ so that $E_{i}\subset {E}$ and ${E} = \cup_{i=1,...,n} E_{i}$. The same holds for ${R}$ when several context relations come together such as "is relative of", "has business affiliation", "has visited", etc. 

By using formal structures within the graph, we are implicitly using the model of a labeled property graph, see \cite{rodriguez2012graph} and \cite{rodriguez2010constructions}. Here, nodes and edges form a heterogeneous set. Nodes and edges can be identified by using a single or multiple labels, for example using $\lambda:E\rightarrow \Sigma$, where $\Sigma$ denotes a set of labels. We need to mention that both concepts are equivalent, since graph databases use the concept of labeled property graphs.

Here, our experimental setting is -- without loss of generality -- settled in social network analysis (SNA). It is quite obvious that a social network containing actors may easily be extended with other data, for example spacial data (e.g. locations, rooms, towns, countries), or social groups (e.g. companies, clubs), or any other information (e.g. information data about actors). 
Once a social network is built, we may start to ask questions like ``How many friends does actor $X$ have?'' or ``To how many groups does actor $Y$ belong?''. The mathematical formulation of these questions would be ``What is the degree of node $X$?'' and ``How many communities $C_i$ can be found such that $Y\in C_i$?''.  The mathematical foundations in this and the following sections are based on  the works of \cite{diestel2012graphentheorie} and  \cite{matouvsek2007diskrete} unless otherwise noted.

In general, we define a \emph{Graph} $G=(V,E)$ with a set of edges or vertices $V$ -- these are actors, locations or any other nodes in the network -- and edges $E$, which describe the relations between nodes. The number of nodes $|V|$ is usually denoted with $n$. Given two nodes $s=$Simon and $j=$Jerusalem we may add an edge or relation $(s,j)$ between both describing for example, that Simon is or was in Jerusalem. Then we say $s$ and $j$ are \emph{connected} or they are \emph{neighbors}. The \emph{neighborhood} of a vertice $v$ is denoted with $N(v)$ and describes all nodes connected to $v$. If we are interested in the size of this neighborhood we calculate the node \emph{degree} given by $deg(v)=|N(v)|$. 

The neighborhood thus gives information about the connectedness of an actor in the network. This can be useful to illustrate the direct influence of an actor within the complete network, especially for actors with a high node degree. But it is obvious that the amount of relations does not necessarily give a good idea on their quality or how we could use these relations. While the node degree is often used as a measure to create random graphs, it is in general not a good measure in order to analyze particular actors in networks, see \cite{Jackson2010}. 

Nevertheless, the \emph{degree centrality} for a node $v\in V$ is given by
\[dc(v)=\frac{deg(v)}{n-1}\]
The output value ranges between 0 and 1 and gives a reference to the direct connections. As discussed, it omits all indirect relations and in particular the node's position in the network.

\begin{definition}[Scale-Free Network]
A network is scale-free if the fraction of nodes with degree $ k $ follows a power law $k^{-\alpha}$, where $\alpha > 1$. 
\end{definition}
\begin{definition}[Small World Network \cite{watts1999networks}]
Let $G=(V,E)$ be a connected graph with $n$ nodes and average node degree $k$. 
Then $G$ is a small-world network if $k\ll n$ and $k\gg 1$.
\end{definition}

\begin{figure}[t]
\centering
\includegraphics[width=0.45\textwidth ]{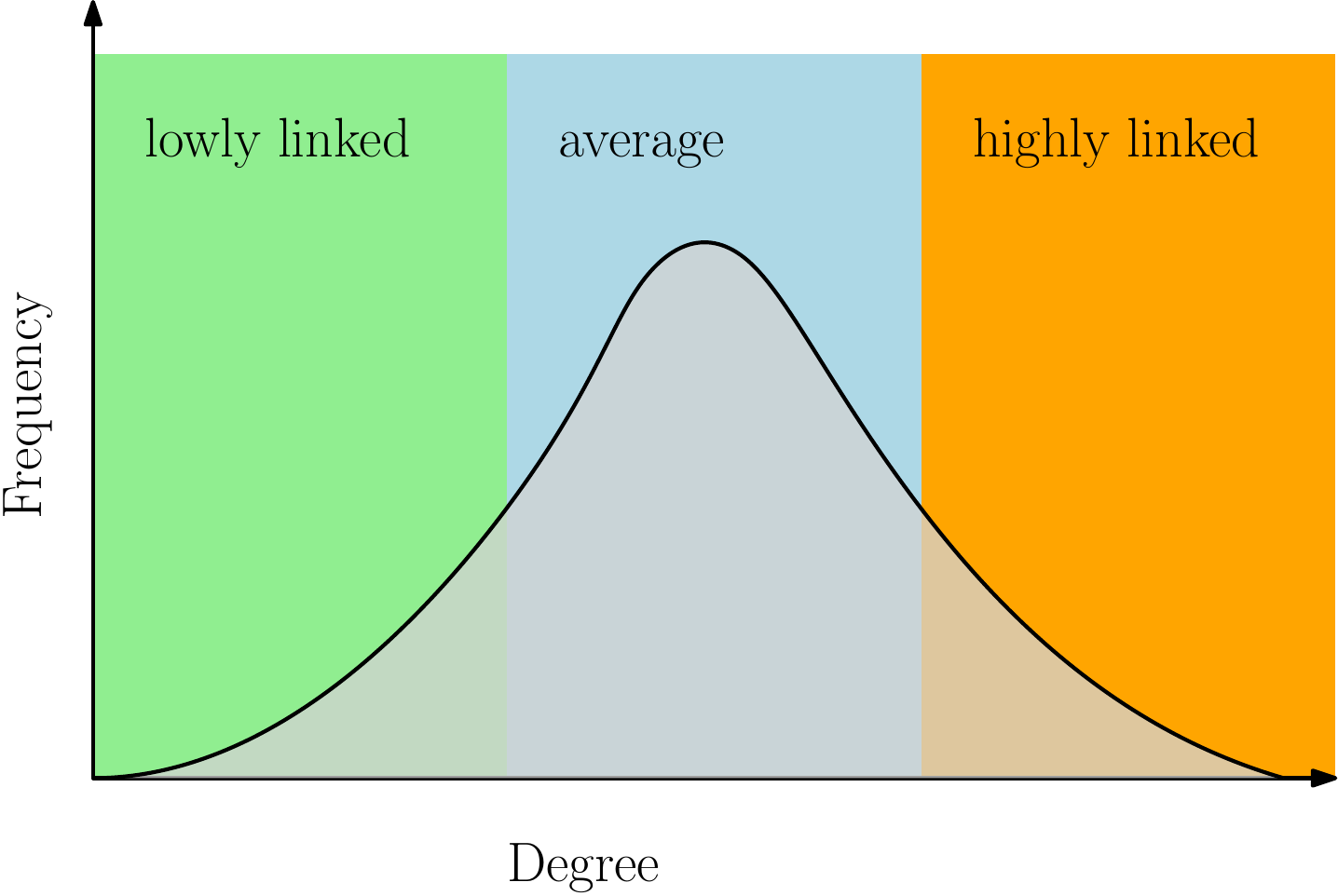} \quad
\includegraphics[width=0.45\textwidth ]{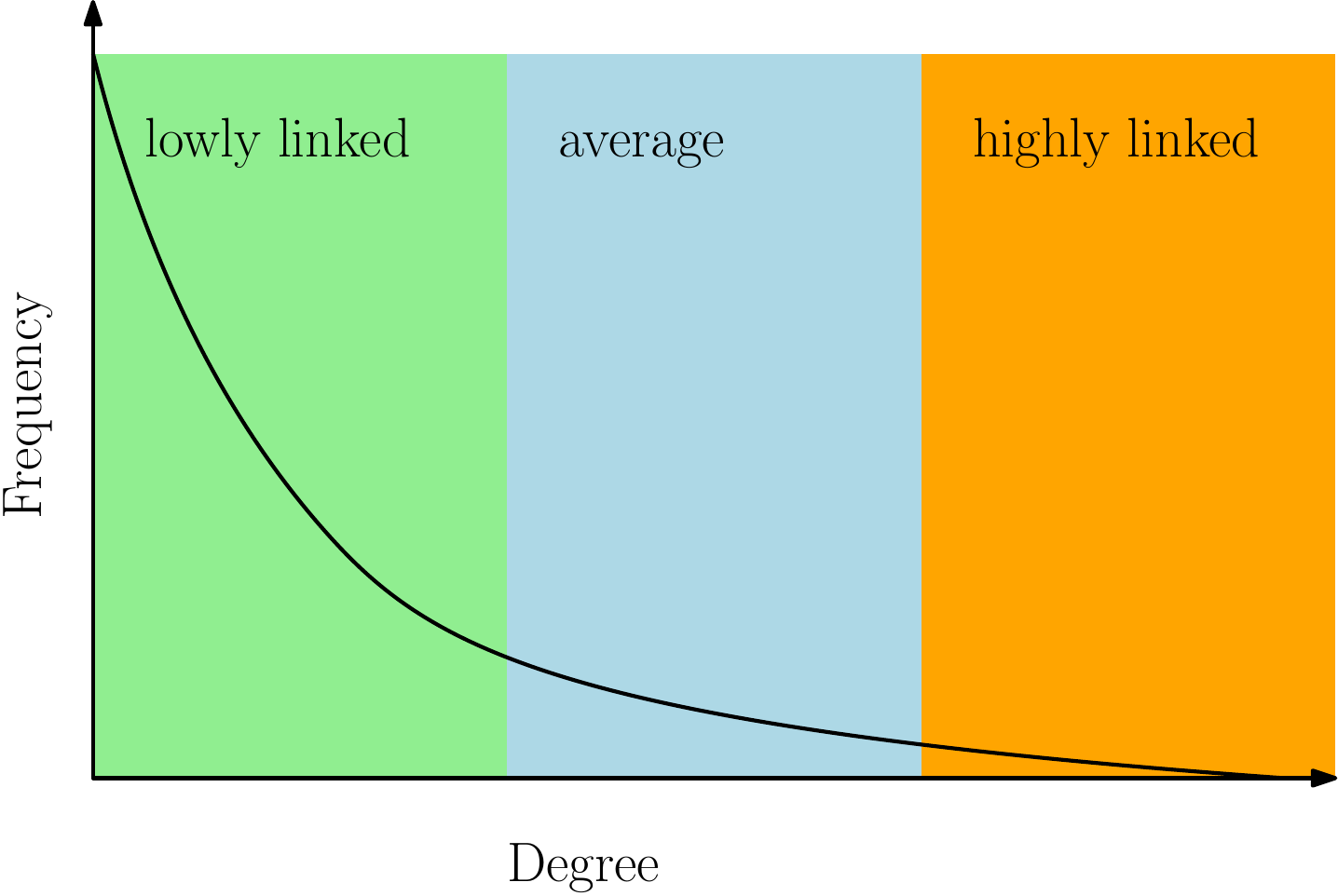}
\caption{Top: In random networks the degree distribution follows a given random distribution. Here, most nodes are average linked and an equal number of nodes is lowly and highly linked. Bottom: Real networks often follow other or even no standard random distribution. Here, a scale-free distribution is shown: Most nodes are lowly linked whereas only very few notes are highly linked.}\label{abb:exl:dist12}
\end{figure}

In any case, the \emph{degree distribution} provides us with information about the network structure since we can distinguish between sparsely and densely connected networks. While \cite{Jackson2010} suggests statistical analysis to compute the correlation between attributes of the network and the density of nodes, this will not work for the small networks and the missing statistical values. In any case, although scale-free networks are not an universal characteristic for real-world networks, we might use this approach to get a first overview about the network itself. Random graphs, like the Erdős–Rényi networks, follow a Poisson distribution. Scale-free networks, inspired by real-world social networks, follow a power law. 
See Figure \ref{abb:exl:dist12} for two examples of a random graph and a more common distribution in real word networks.

We will now discuss one more property to evaluate nodes and their position in the networks. 
These properties can be used to calculate statistical parameters, so-called \emph{centrality measures}, cf. \cite{freeman1978centrality} and \cite{carrington2005models}. They answer the question ``Which nodes in this network are particularly significant or important?''.

\emph{Betweenness} analyzes critical connections between nodes and thus gives an indication of individuals that can change the flow of information in a network. This measure is based on paths in a network:

\begin{quote}
Much of the interest in networked relationships comes from the fact that individual
nodes benefit (or suffer) from indirect relationships. Friends might provide access
to favors from their friends, and information might spread through the links of
a network.\cite{Jackson2010}
\end{quote}
A \emph{path} $p$  in a graph $G=(V,E)$ is a set of vertices $v_1,...,v_t$, $t \in \mathbb{N}$, for example written as
\[p=[v_1,...,v_t],\] 
where $(v_i,v_{i+1})\in E$ for $i\in\{1,\ldots,t-1\}$. 
The length $|p|$ of the path $p$ is the total number of edges -- not nodes. Thus $|p|=t-1$. 
The path $p$ links the starting node $v_1$ and an ending node $v_t$. 
In a path, no crossings are allowed, thus $v_i\neq v_j$ for all $i,j\in \{1,...,t\}$. 
If all properties of a path are met except that the beginning and the end vertex are the same -- that is, $v_1 = v_t$ -- 
we denote this set as a \textit{circle}.

\emph{Betweenness centrality} was first introduced by \cite{freeman1977set}\footnote{Initially introduced for symmetric relations -- undirected graphs -- it was extended to asymetric relations -- directed graphs -- by \cite{white1994betweenness}.} and considers other indirect connections, see \cite{schweizer1996muster}. Given a node $v$, it calculates all shortest paths in a network $P_v(k,j)$ for all beginning and ending nodes $k,j\in V$ that pass through $v$. If $P(k,j)$ denotes the total number of paths between $k$ and $j$, the importance of $v$ is given by the ratio of both values. Thus the betweenness centrality according to \cite{Jackson2010} is given by

\[bc(v)= \sum_{k\neq j, v\neq k, v \neq j} \frac{P_v(k,j)}{P(k,j)} \cdot \frac{2}{(n-1)(n-2)},\] 

where $ n $ denotes the number of the vertices in the graph. 
This parameter allows an analysis of the critical links and how often a node lies on such a path. This centrality measure thus answers the questions whether a node can change the flow of information in a network or whether it is a bridge between other nodes, see \cite{schweizer1996muster}.

While betweenness assumes network flows to be like packages flowing from a starting point to a destination, other measures consider multiple paths: For example, the so-called \emph{eigenvector centrality} -- introduced by \cite{bonacich1972factoring} -- measures the location of directly neighboring nodes in the network. For the 
eigenvector centrality, we ``count walks, which assume that trajectories can not only be circuitous, but also revisit nodes and lines multiple times along the way.''\cite{borgatti2005centrality} This measure not only classifies the direct possibility to influence neighbors, but also ranks the indirect possibility to influence the whole network. For a detailed mathematical background we refer to~\cite{Jackson2010}.

Less popular measures are Katz prestige, and Bonacich’s measure, see \cite{Jackson2010}. It has been shown that these measures are closely related, see \cite{ditsworth2019community}.

\section{Method}

\begin{figure}[t] %
	\centering
	\includegraphics[width=0.49\textwidth]{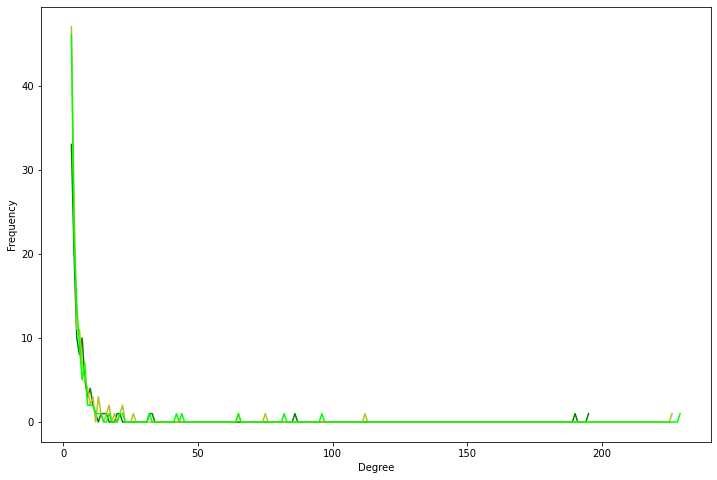}
	\caption{Frequency of nodes with a given degree for three random Scale-Free Networks with $n=150$ nodes. }
	\label{img:fd1}
\end{figure}

We evaluate the degree centrality and betweenness centrality on random graphs. First, we consider Scale-Free Networks with $n$ nodes, 
see~\cite{Jackson2010}. Moreover, \cite{bollobas2003directed} introduced a widely used graph model with 
three random parameters $\alpha+\beta+\gamma=1$. 
These values define probabilities and thus define attachment rules to add new vertices between either existing or new nodes. This model allows loops and multiple edges, where a loop denotes one edge where the endvertices are identical, and multiple edges denote a finite number of edges 
that share the same endvertices. Thus, we convert the random graphs to undirected graphs. For testing purpose, we scale the number of nodes $n$ and use 
$\alpha=0.41$, $\beta=0.54$, and $\gamma=0.05$.
We chose this random graph model since it is generic and feasible for computer simulations for measuring and evaluation purposes, 
see~\cite{bollobas2003mathematical,kivela2014multilayer}. 

Figure \ref{img:fd1} shows the frequency of nodes (y-axis) with a particular degree (x-axis) for three random networks with $n=150$ nodes. 
Compared to Figure~\ref{img:fd1}, Figure~\ref{abb:exl:dist12} clearly shows the scale-free distribution, in which many nodes have a small degree and only few 
nodes have a very large degree: most nodes are hence lowly linked. Thus these small-degree nodes lead to a few communities which are highly connected.

\begin{figure}[t] %
	\centering
	\includegraphics[width=0.49\textwidth]{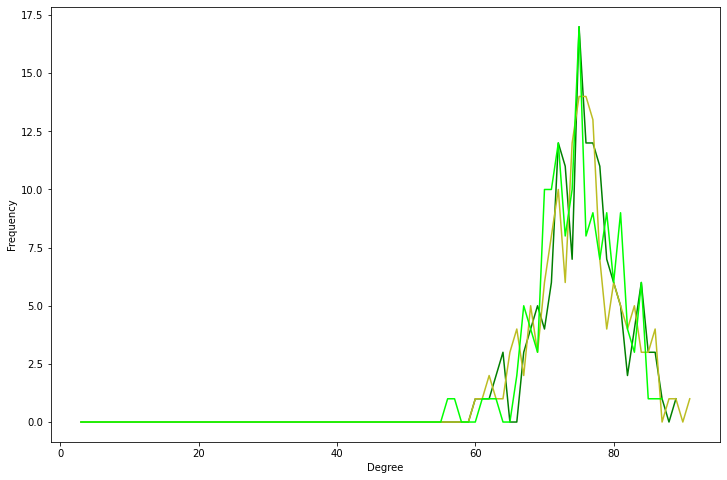}
	\caption{Frequency of nodes with a given degree for three Newman-Watts-Strogatz small-world random graph with $n=500$ nodes. }
	\label{img:fd2}
\end{figure}

The second random graph uses a fixed degree distribution and is widely known as Newman-Watts-Strogatz small-world random graph \cite{NEWMAN1999341}. The algorithm to create such as graph takes a number of nodes $n$, the number of $k$ nearest neighbors that form a ring topology and the probability  $p$ for adding a new edge. A small-world graph contains only small average paths and thus has a small diameter, see \cite{Jackson2010}. Some studies like \cite{aarstad2013ways} study the relation between scale-free and small-world networks, in particular the relationship between the average path length and local clusterings. In general, it is possible to generate scale-free networks with small-world attributes, see \cite{klemm2002growing}.

Figure~\ref{img:fd2} shows the frequency of nodes with a given degree for three random networks with $n=500$ nodes. Compared to Figure~\ref{abb:exl:dist12}, 
Figure~\ref{img:fd2} clearly shows the Poisson distribution with many nodes having an average degree. Together with Figure~\ref{img:fd1} it also illustrates the ``long tail'' of the scale-free distribution, see \cite{Jackson2010}. 

We will now evaluate how graph structures and in particular measures change when additional information are stored in extra layers. 
We partition a graph into an uncolored part that contains the `original' data and into a part with blue nodes in which novel `extra' data stored. 
These blue nodes simulate one or more new layers in the knowledge graph.
Thus, given a random graph $G=(V,E)$, a next step comprises a probability $p_b$ for blue nodes which leads to a graph $G$ with blue nodes $B\subset V$. 
First, we compute the centrality measures for all nodes in $G_B = (V\setminus B, E)$ and then for all nodes in $G$ but limit the output to all nodes in $B$. 
Thus, we have two vectors $c_1, c_2\in \mathbb{R}^n$ where $n$ is the number of nodes in $V\setminus B$. 
We denote $c_i$ by $c_i = \left( c_i^1, c_i^2, c_i^3, ...\right)$.
 
While comparing two vectors, we are interested in two values. 
The first one is the total number of misordered elements, that is, 
the total number of positions on which the elements differ from each other. 
The second value that we compute in order to compare two vectors 
is the number of moved elements. 
For this we count those elements that have a different predecessor and / or successor in the first vector compared to the second one.   

\begin{exl}\label{ex:errormeasures}
Let $c_1 = [1,2,3,4,5]$, $c_2=[5,3,2,1,4]$ and $c_3=[1,5,2,3,4]$. If $c_1$ is the original ordering, we see that $c_2$ has a totally different order. In $c_3$ the entry $5$ is moved, but the rest of the list is unchanged, although still 4 elements are on the wrong location. Hence, the number of misordered elements in $c_1$ compared to $c_2$ is 5. The number of moved elements is 5 and 1. %
\end{exl}
To identify both errors, we first define function $e$:

\[e(i,c_1, c_2)=\begin{cases}
    0 & c_{1}^{i}=c_{2}^{i}\\
    1 & c_{1}^{i}\neq c_{2}^{i}
\end{cases}\]

That is, $e(i,j,c_1,c_2)=1$ if the element on the $i$th position of $c_1$ 
differs from the element on the $j$th position in $c_2$. 
To shorten notation, we write $e(i,c_1,c_2)$ whenever $i=j$.

Let $x$ be an element contained in every $c_u$, $u \in \mathbb{N}$. %
Then $ p(x,c_u) $ denotes the predecessor of element $ x $ in $c_u$ 
and $s(x,c_u)$ denotes the successor of $x$ in $c_u$.  
If $x$ is the first element in $c_u$, then $p(x,c_u)=\emptyset$. 
If $x$ is the last element of $c_u$, then $s(x,c_u)=\emptyset$. 
With these definitions, 
we define $e_N$: 

\[e_N(x,c_1, c_2)=\begin{cases}
    1 	& \text{if } p(x,c_1)=\emptyset \text{ and } s(x,c_1) \not = s(x,c_2),\\
        & \text{or } s(x,c_1)=\emptyset \text{ and } p(x,c_1) \not = p(x,c_2),\\
        & \text{or } s(x,c_1) \not = s(x,c_2) \text{ and } p(x,c_1) \not = p(x,c_2),\\
    1/2	& \text{if } s(x,c_1) \not = s(x,c_2) \text{ and } p(x,c_1) = p(x,c_2),\\
        & \text{or } s(x,c_1) = s(x,c_2) \text{ and } p(x,c_1) \not = p(x,c_2),\\
    0	& otherwise. 
\end{cases}\]

In other words, 
we consider the predecessor of an element in $c_1$ and 
check if this element is still a predecessor of this element in $c_2$, 
and analyse analoguously the successor of an element.

With this, we define two error measures $\epsilon$ and $\epsilon_N$:

\[\epsilon(c_1, c_2)= \sum_{i=1}^n e(i,c_1, c_2)\]

\[\epsilon_N(c_1, c_2)= \sum_{x \in c_1} e_N(x,c_1, c_2)\]

\begin{exl}
	Let's reconsider Example~\ref{ex:errormeasures}: 
	Recall that $c_1 = [1,2,3,4,5]$, $c_2=[5,3,2,1,4]$ and $c_3=[1,5,2,3,4]$. 
	Then, $\epsilon(c_1,c_2)=5 \text{ and } \epsilon_{N}(c_1,c_2)=5.$  
	Moreover, $\epsilon(c_1,c_3)=4$ and 
	$\epsilon_{N}(c_1,c_3)=2.5$.
\end{exl}

We will now analyze different scenarios to evaluate the impact of additional blue nodes on a scale-free and a small-world network.

\section{Results}

\subsection{Degree Centrality}

\begin{figure*}[t] %
	\centering
	\includegraphics[width=0.69\textwidth]{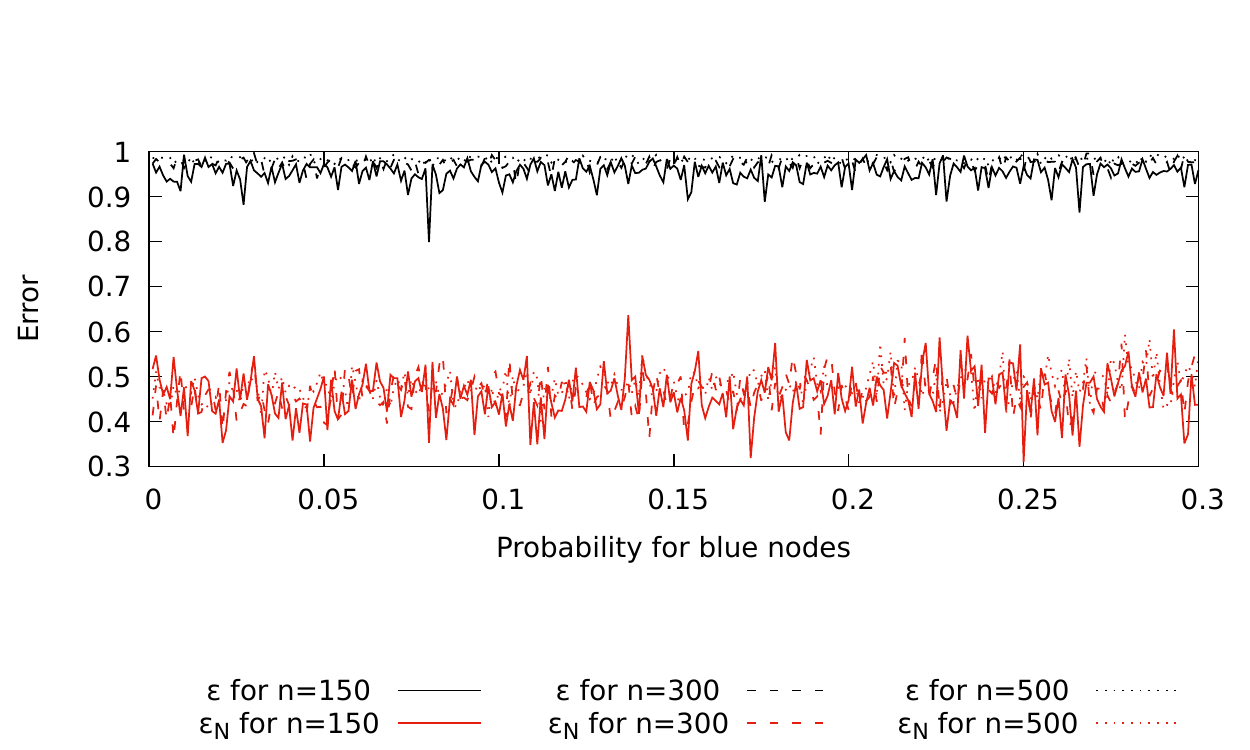}
	\caption{Degree Centrality errors for scale-free random graphs ($n=150$, $n=300$ and $n=500$) for different values of $p_B$ between 0 and~0.3. }

	\label{img:dc-error-1}

	\centering
	\includegraphics[width=0.69\textwidth]{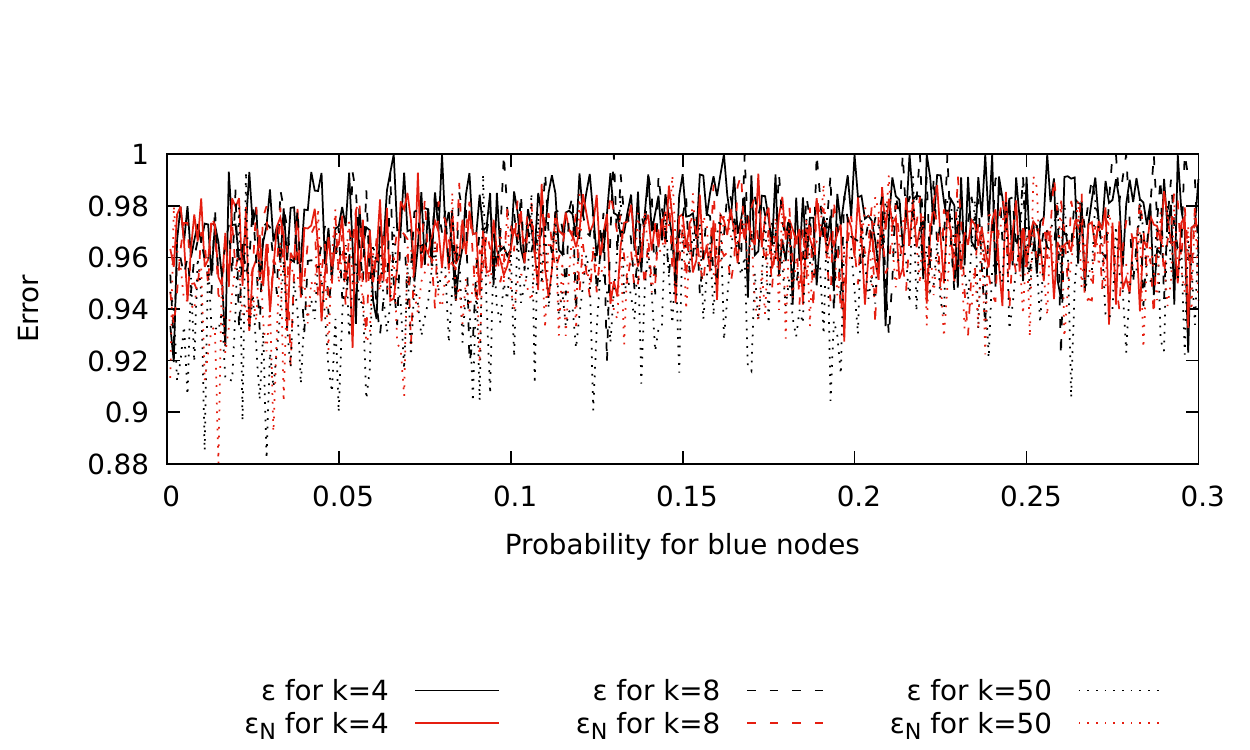}
	\caption{Degree Centrality errors for Newman-Watts-Strogatz small-world random graph ($n=150$, $k\in\{4,8,50\}$) for different values of $p_B$ between 0 and~0.3. }

	\label{img:dc-error-2}
\end{figure*}

The Degree Centrality was evaluated with errors $\epsilon$ and $\epsilon_N$ for scale-free random graphs ($n=150$, $n=300$ and $n=500$, see Figure \ref{img:dc-error-1}) and Newman-Watts-Strogatz small-world random graphs ($n=150$, $k\in\{4,8,50\}$, see Figure \ref{img:dc-error-2}). The mean values are given in Table \ref{tab:1}.

\begin{table}
\centering
\begin{tabular}{|c|c|c|c|c|c|c|}
\hline 
 & $\epsilon$ & $\epsilon_{N}$ & $\epsilon$ & $\epsilon_{N}$ & $\epsilon$ & $\epsilon_{N}$\tabularnewline
\hline 
Scale-Free & \multicolumn{2}{c|}{$n=150$} & \multicolumn{2}{c|}{$n=300$} & \multicolumn{2}{c|}{$n=500$}\tabularnewline
\hline 
Mean & 0.95 & 0.46 & 0.97 & 0.47 & 0.98 & 0.48\tabularnewline
\hline 
Small-World & \multicolumn{2}{c|}{$k=4$} & \multicolumn{2}{c|}{$k=8$} & \multicolumn{2}{c|}{$k=50$}\tabularnewline
\hline 
Mean & 0.97 & 0.97 & 0.97 & 0.96 & 0.95 & 0.96\tabularnewline
\hline 
\end{tabular}
\vspace*{.1cm}
\caption{Mean values for Degree Centrality errors.}\label{tab:1}
\end{table}
Here, we see that the Small-World graph has a very high error rate for both $\epsilon$ and $\epsilon_N$ even for small $p_B$. In particular, the values are rather constant, no matter what value was chosen. In addition, the graph topology for different values of $k$ has only very little impact on the error rate. Thus, even small changes in the graph structure (a very small value for $p_B$) have a great impact on the degree centrality. Since Small-World graphs have a high level of local clustering, the random exclusion of blue nodes will most likely effect not only one cluster, but also other clusters. This changes not only the position, but also the ordering of node degrees. 

A different scenario occurs when considering Scale-Free graphs. Again we see a very high error rate for $\epsilon$, even for small $p_B$. The values for $\epsilon_N$ are usually near to $.5$ (mean values 0.46, 0.47, 0.48). Neither the graph size $n$ nor the value for $p_B$ has an impact on these errors. Here, we see the scale-free distribution: the blue nodes do change the position of the degree centrality, but while they also change the ordering within clusters, they do not affect the complete ordering due to the longer distance between nodes.

\subsection{Betweenness Centrality}

\begin{figure*}[t] %
	\centering
	\includegraphics[width=0.69\textwidth]{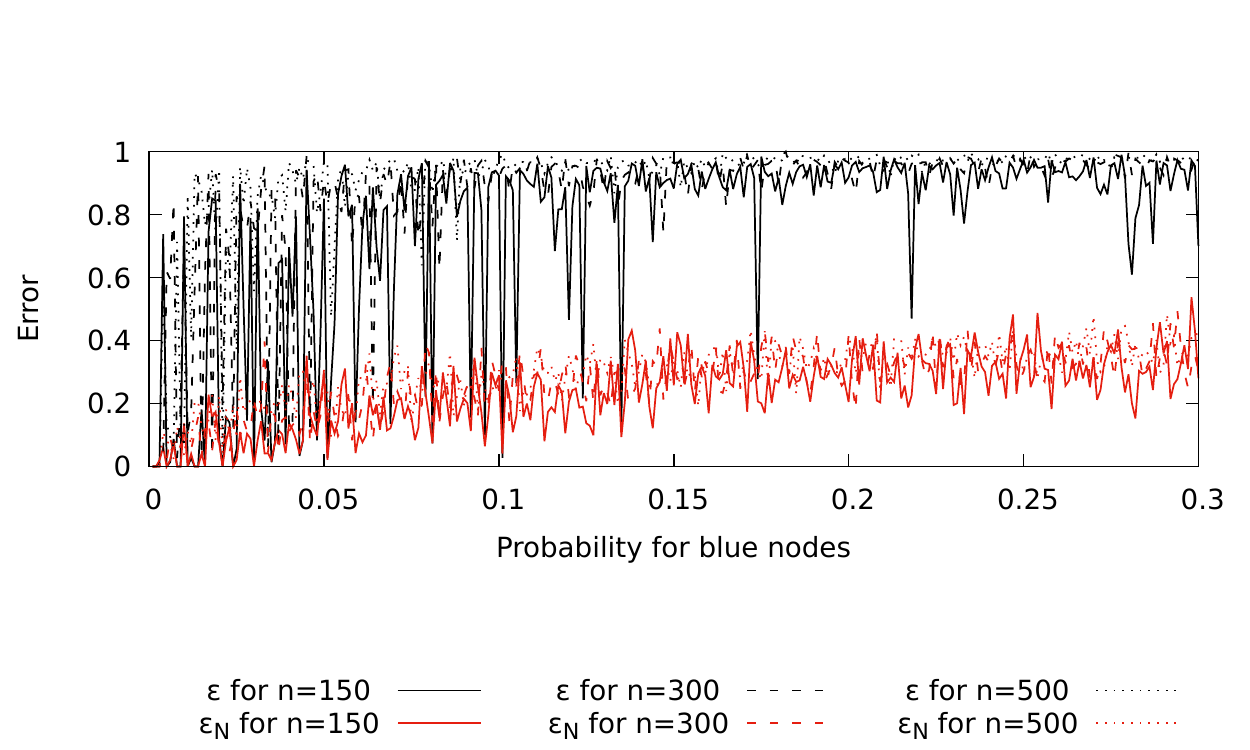}
	\caption{Betweenness Centrality errors for scale-free random graphs ($n=150$, $n=300$ and $n=500$) for different values of $p_B$ between 0 and 0.3. }

	\label{img:bc-error-1}

	\centering
	\includegraphics[width=0.69\textwidth]{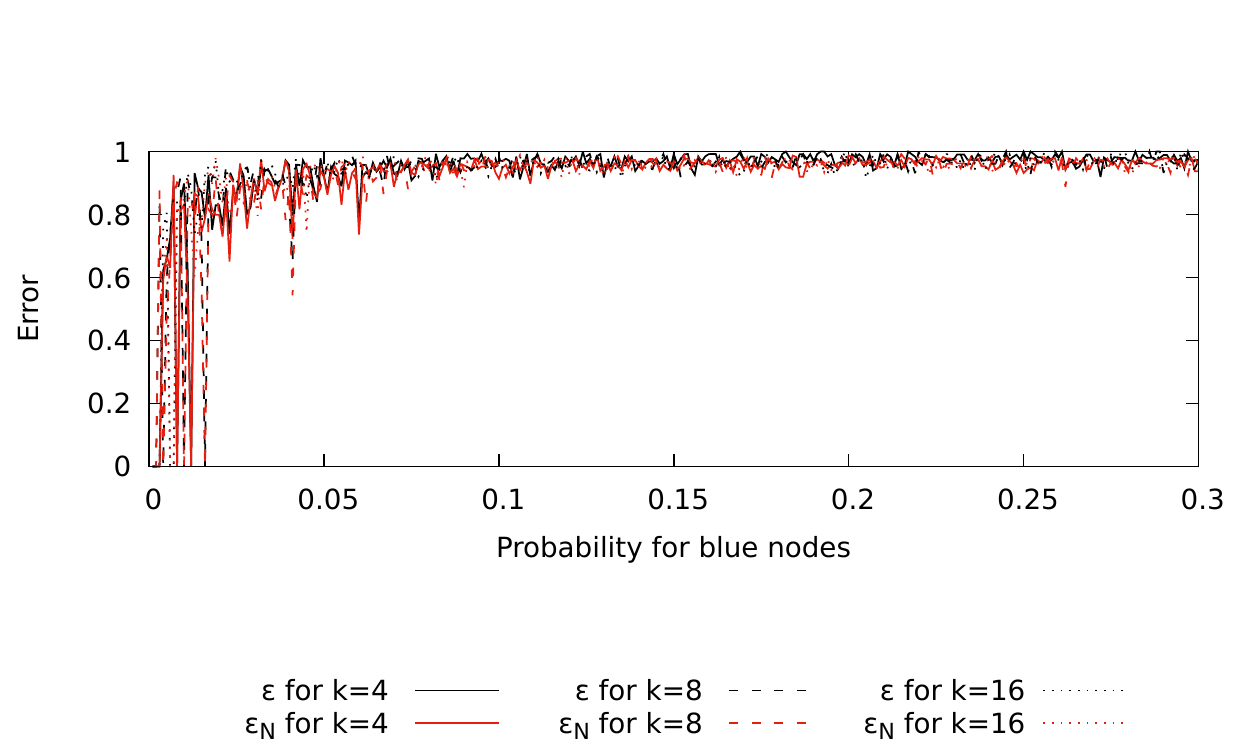}
	\caption{Betweenness Centrality errors for Newman-Watts-Strogatz small-world random graph ($n=150$, $k\in\{4,8,50\}$) for different values of $p_B$ between 0 and 0.3. }

	\label{img:bc-error-2}
\end{figure*}

The Betweenness Centrality was evaluated with errors $\epsilon$ and $\epsilon_N$ for scale-free random graphs ($n=150$, $n=300$ and $n=500$, see Figure \ref{img:bc-error-1}) and Newman-Watts-Strogatz small-world random graphs ($n=150$, $k\in\{4,8,50\}$, see Figure \ref{img:bc-error-2}). The mean values are given in Table \ref{tab:2}.

\begin{table}
\centering
\begin{tabular}{|c|c|c|c|c|c|c|}
\hline 
 & $\epsilon$ & $\epsilon_{N}$ & $\epsilon$ & $\epsilon_{N}$ & $\epsilon$ & $\epsilon_{N}$\tabularnewline
\hline 
Scale-Free & \multicolumn{2}{c|}{$n=150$} & \multicolumn{2}{c|}{$n=300$} & \multicolumn{2}{c|}{$n=500$}\tabularnewline
\hline 
Mean & 0.77 & 0.23 & 0.87 & 0.27 & 0.91 & 0.29\tabularnewline
\hline 
Small-World & \multicolumn{2}{c|}{$k=4$} & \multicolumn{2}{c|}{$k=8$} & \multicolumn{2}{c|}{$k=50$}\tabularnewline
\hline 
Mean & 0.94 & 0.92 & 0.94 & 0.92 & 0.94 & 0.93\tabularnewline
\hline 
\end{tabular}
\vspace*{.1cm}
\caption{Mean values for Betweenness Centrality errors.}\label{tab:2}
\end{table}

Betweenness centrality (see Figure \ref{img:bc-error-1}) in scale-free graphs is very much influenced by the choice for $p_B$. Again, the total error $\epsilon$ becomes very high although there are several outliers. More interesting is again the ordering error $\epsilon_N$: although the error increases with a rising value of $p_B$, it remains very low. Again, the number of nodes $n$ has only very little impact on the error measures.

Here, again, the Small-World graph has a very high error rate for both $\epsilon$ and $\epsilon_N$ although not for very small $p_B$, see Figure \ref{img:bc-error-2}. 
In particular, we may find a boundary $p'_B$ so that the values are rather constant for $p_B>p'_B$. Again, the graph topology for different values of $k$ has only very little impact on the error rate. Thus, even small changes in the graph structure (a very small value for $p_B$) have a great impact on the betweenness centrality. Thus, the random choice of blue nodes again destroys the structures of local clustering which will most likely effect not only one cluster, but also other clusters.

\section{Discussion and Outlook}

This paper investigates the impact of a multiple layers on centrality measures compared to single-purpose graph. We presented an experimental environment to evaluate two different centrality measures -- degree and betweenness centrality -- on random graphs inspired by social network analysis: small-world and scale-free networks. The result clearly shows  that the graph structures and topology have a great impact on its robustness for additional data stored. In particular, we could identify nodes with a high node degree and closely connected communities or clusters as problematic for reordering the centrality measures. Thus, we could show that small-world networks are rather less robust than scale-free networks. 

Although the experimental analysis of random graphs allows us to make some basic observations we could also present some very preliminary error approximations. %
We need to mention that a lot of research needs to be done in this field, because we only considered degree and betweenness centrality. 
In particular, we can identify the following questions for further research: Is it possible to find good error approximations for larger sets of blue nodes $B$? How do $\epsilon$ and $\epsilon_N$ behave on any given node $v\in B\subset V$ with a node degree $d(v)=m$? What are (other) graph structures that have a great impact on the stability of networks for degree, betweenness and other centralities? 

To sum up, it is valid %
to extend single-purpose networks with data from other sources. In particular, we considered random social networks as a basis. Thus, extending social networks with other information layers is possible, although it will change the behavior of measurements like network centrality. The effect highly depends on the given graph structure. More interdisciplinary research is needed to investigate the impact on real-world data within the context of humanities.

\bibliographystyle{IEEEtran}
\bibliography{lit}

\end{document}